\documentclass{aa} 
\usepackage{graphicx}
\begin{document}
   \title{Multi-object spectroscopy of low redshift EIS clusters. 
          I.\thanks{Based on observations made with
          the Danish1.5-m telescope at ESO, La Silla, Chile}}

   \author{L. Hansen
           \and
           L.F. Olsen
           \and
           H.E. J{\o}rgensen
          }

   \offprints{L. Hansen}

   \institute{Copenhagen University Astronomical Observatory,
              Juliane Maries Vej 30,
              DK-2100 Copenhagen, Denmark
             }

   \date{Received .....; accepted .....}

   \abstract{ We report the results of the first multi-object
spectroscopic observations at the Danish 1.54m telescope at
La~Silla, Chile. Observations of five cluster candidates from the ESO Imaging
Survey Cluster Candidate Catalog are described. From these
observations we confirm the reality of the five clusters with measured
redshifts of $0.11\leq z\leq0.35$. We estimate velocity dispersions in
the range 294--621\,km/s indicating rather poor clusters. This, and the
measured cluster redshifts are consistent with the results of the matched
filter procedure applied to produce the Cluster Candidate Catalog.
   }

   \maketitle
%

\section{Introduction}

The evolution in the properties of galaxy clusters and their
constituent galaxies is an important issue of contemporary cosmology
and astrophysics.  The demand for large samples of clusters at various
redshifts has prompted major systematic efforts to provide catalogues
of distant clusters (for references see e.g. Holden et
al. \cite{Holden}). Olsen et al.  (\cite{Olsen99a}, \cite{Olsen99b})
and Scodeggio et al. (\cite{Scodeggio99}) presented the ESO Imaging
Survey (EIS) Cluster Candidate Catalog containing a total of 302
cluster candidates with estimated redshifts of $0.2 \le z \le
1.3$. Their work is based on the EIS I-band images covering a total
area of 17 square degrees. The candidates were selected via a
matched filter algorithm akin to the one introduced by Postman et
al. (\cite{Postman96}).

In order to confirm the reality of the EIS cluster candidates and
determine membership of the clusters Ramella et al. (\cite{Ramella})
have initiated multi-object spectroscopy (MOS) with the 3.6m ESO
telescope at La~Silla, Chile, for the $0.5 \le z \le 0.7$ part of the
sample. Furthermore, the spectroscopic investigation of the high
redshift candidates ($0.8 \le z \le 1.3$) using the VLT has been
initiated by Benoist et al. (\cite{Benoist}). They report the
detection of bound systems in 3 candidate fields. One of the systems
coincides with an X-ray source recently detected by XMM 
(Neumann et al. \cite{Neumann}). 

Here we present the first results for the low redshift part ($0.2 \le
z \le 0.4$) based on MOS obtained at the Danish 1.54m telescope at
La~Silla. The data were taken as test observations aiming at
implementing MOS at this telescope. The observations were performed during a
run in February/March 2001 when moon or weather prevented the
photometric main program, and exposures through the eight available slit 
masks covering five cluster candidates (Table~\ref{ObsClusters})
were secured. Unfortunately
an arrangement suitable for proper flat fielding of the multi-slit images
was not yet installed. Flat fields against a screen fixed in the dome
were obtained instead, but as described below this procedure did not
optimize the quality of the final spectra. Nevertheless, seven masks
out of eight had spectra adequate for redshift determination
(Table~~\ref{plates}).


\section{Observations}

The Danish Faint Object Spectrograph and Camera (DFOSC) mounted at the
Danish 1.54m telescope is well suited for MOS on galaxy clusters of
moderate redshift. Multi-slit masks can be placed in the slit wheel
which is easily accessible, and the field is as large as $13 \farcm 7
\times 13 \farcm 7$ matching the extent of the moderate redshift
clusters. A challenge is the small scale in the telescope focal
plane. The slit width used is $2 \arcsec$ or only $128\,\mu$m. The length of 
the slits varied depending on the extent of the galaxies.

The multi-slits are milled in thin (0.85\,mm) plastic sheets of
selected quality in the workshop of the Copenhagen University Astronomical 
Observatory. The positions and lengths of the slits are determined from
pre-imaging frames of the fields using software that converts the
positions on the chip to positions on the slit mask. The
transformations are found using a mask with an accurately known grid of
pinholes. An exposure through this mask is obtained, and the positions
of the images of the pinholes are measured. This procedure has been
carried out for two pinhole-masks (P. Kj{\ae}rgaard, private communication).
The accuracies of the two
transformations are $2.5\,\mu$m and $5.3\,\mu$m, corresponding to $0
\farcs 04$ and $0 \farcs 08$ rms, respectively, which gives an
estimate of the positional accuracy of the slits.  The software also
allows the observer to avoid overlapping spectra by displaying the
computed positions of the spectra on the chip.  For aligning the
slit mask with the sky, holes corresponding to the position of at least
three stars are made.

Alignment of the mask with the sky is carried out using direct
exposures of the cluster field and of the slit mask without
grism. Offset and rotation are computed based on the positions of the
stars and the corresponding holes in the  mask using software
developed by the local staff (Jones et al. \cite{Jones}). The field
and mask are aligned by offset of the telescope, rotation of DFOSC and
possibly fine rotation of the filter wheel. The fine rotation is
necessary, because DFOSC cannot be rotated with the required accuracy.

All spectra were obtained with grism \#4 giving a dispersion of
220\,{\AA}/mm. The final spectra were extracted in the region
3800-7500\,{\AA}, but the useful range depended on the spectrum. The
resolution determined from He-Ne line spectra is found to be
16.6\,{\AA} FWHM.  Total exposure times of minimum 1~hour were aimed
at for each mask, but the images were read out each 20~minutes to
facilitate removal of cosmic ray events in the images. 

For the present observations a sample of 5 cluster candidates
(Table~\ref{ObsClusters}) were selected from Table~1 of Scodeggio et
al.  (\cite{Scodeggio99}). All of these have well developed red
sequences (Olsen \cite{Olsen_thesis}).  The target galaxies were
selected along the red sequence, i.e. cluster galaxies of early
morphological types are favored.  A mask typically contains
$\approx20$ slits, and one or two masks per field covered most of our
candidate objects.

   \begin{table*}
      \caption[]{Observed cluster candidates. The IDs refer to the EIS 
Cluster Candidate Catalogue (Scodeggio et al. (\cite{Scodeggio99}). 
$ z_{\rm mf}$
is the matched filter estimated redshift. ${\rm \Lambda_{cl}}$ is a parameter
of the matched filter method measuring cluster richness. ${\rm N_{R}}$
is an Abell-like richness parameter}
         \label{ObsClusters}
      \center
      \begin{tabular}{l l l l l l l }
            \hline
            Field & ${\rm \alpha_{J2000}}$ & ${\rm \delta_{J2000}}$ & 
            $ z_{\rm mf}$ & ${\rm \Lambda_{cl}}$ & ${\rm N_{R}}$ & mask \\
            \hline
 EIS0946-2029 & 09:46:12.8 & -20:29:49.6 & 0.2 & 59.5 & 44 & 11 \& 12  \\
 EIS0947-2120 & 09:47:06.9 & -21:20:55.6 & 0.2 & 43.4 & 50 & 13 \& 14  \\
 EIS0948-2044 & 09:48:07.9 & -20:44:31.2 & 0.2 & 42.8 & 32 & 15 \& 16  \\
 EIS0949-2101 & 09:49:22.1 & -21:01:47.1 & 0.3 & 37.7 & 57 & 18        \\
 EIS0956-2059 & 09:56:25.2 & -20:59:49.8 & 0.3 & 39.7 & 37 & 17        \\
            \hline
         \end{tabular}
         \end{table*}


%
\begin{figure*}
\begin{center}
  \resizebox{0.5\columnwidth}{!}{\includegraphics{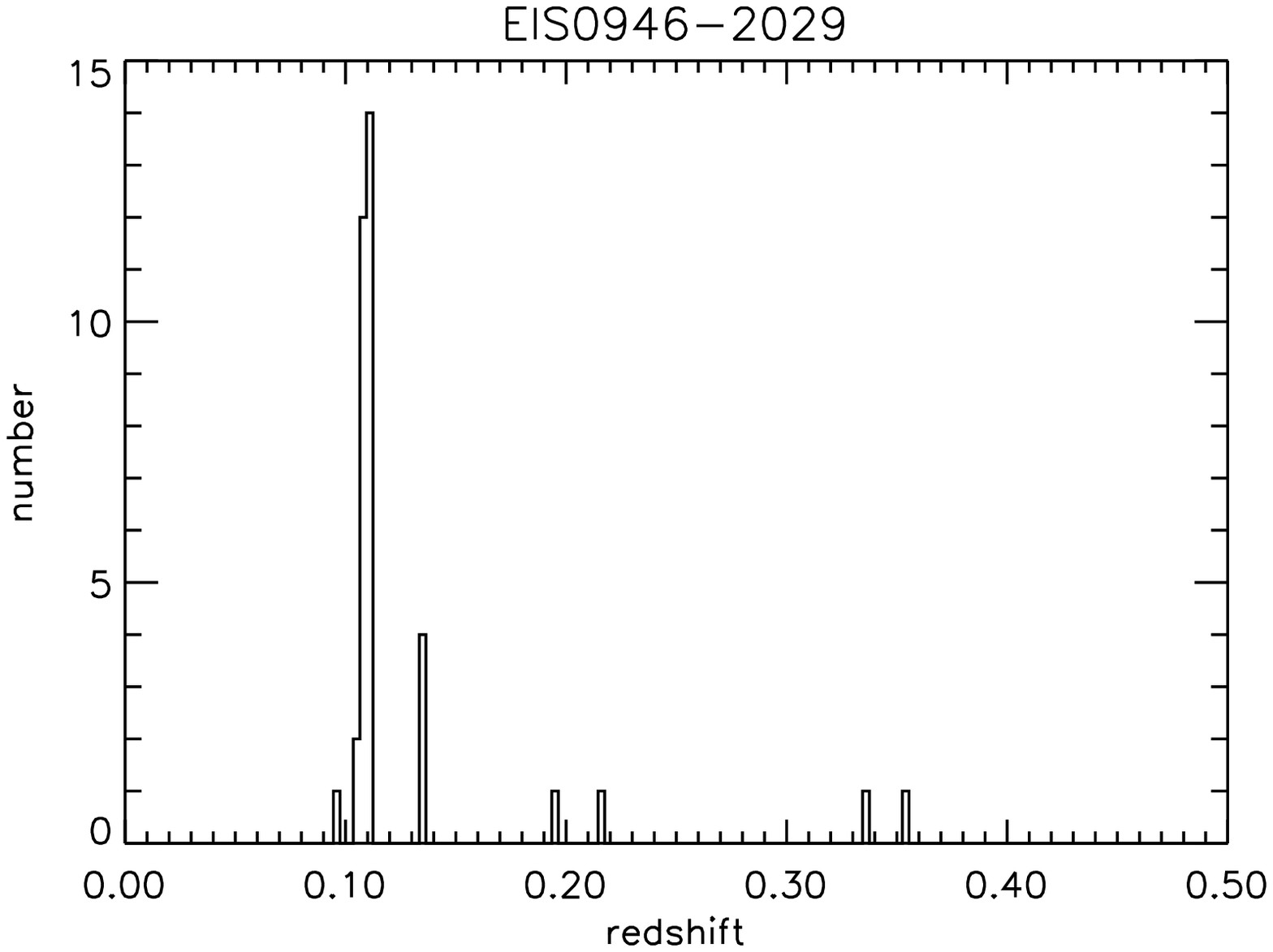}}
  \resizebox{0.5\columnwidth}{!}{\includegraphics{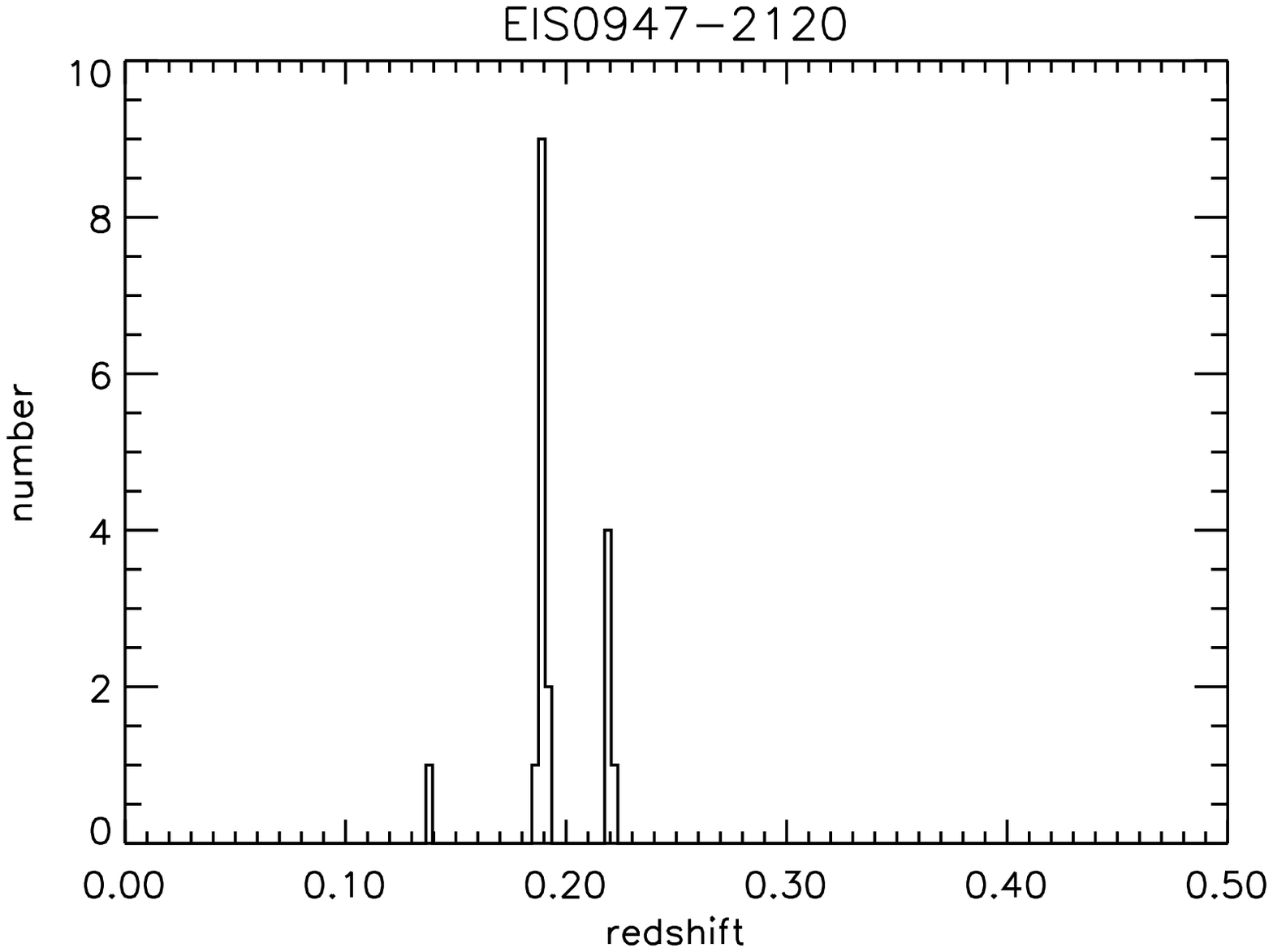}}
  \resizebox{0.5\columnwidth}{!}{\includegraphics{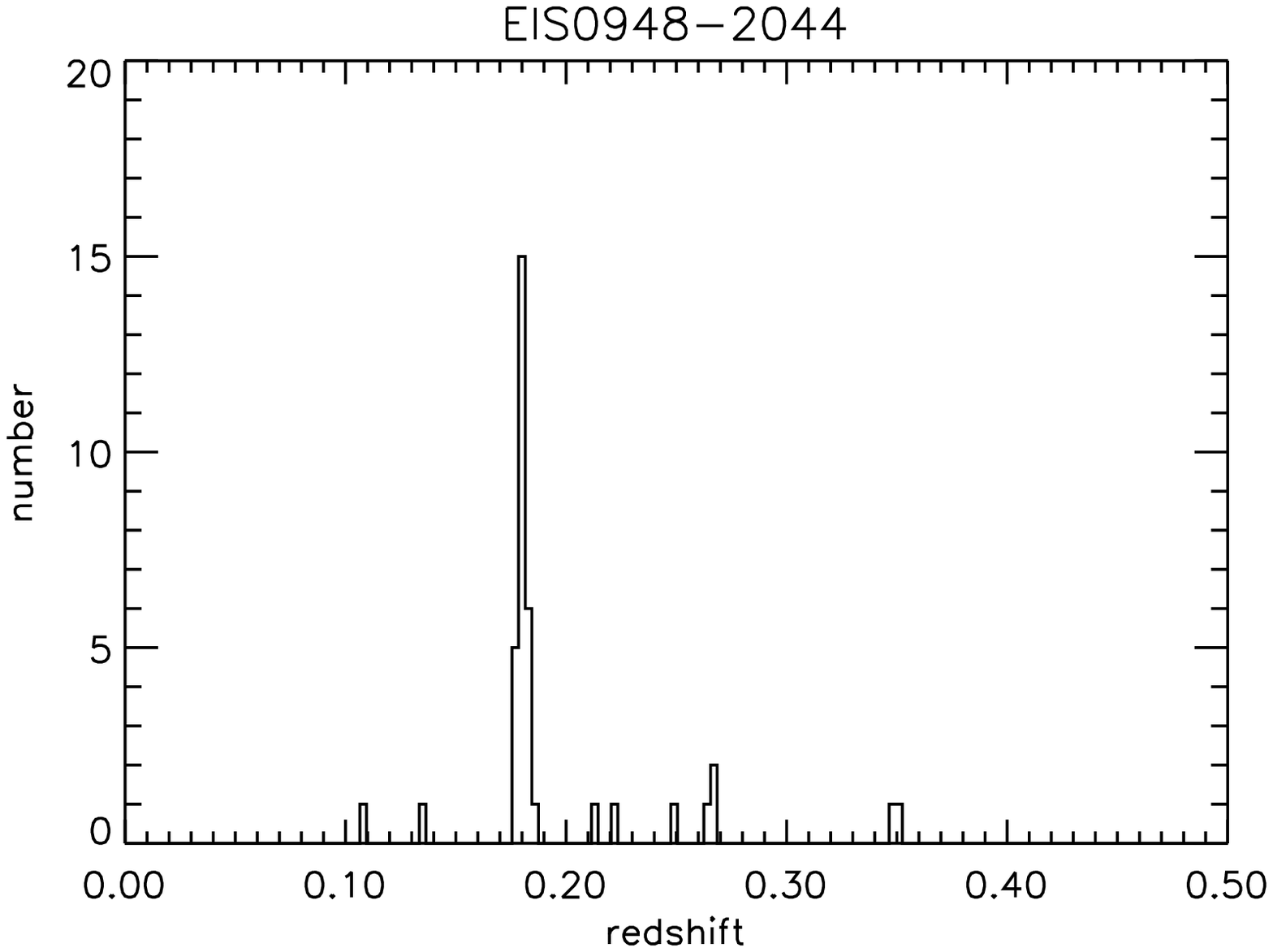}}
  \resizebox{0.5\columnwidth}{!}{\includegraphics{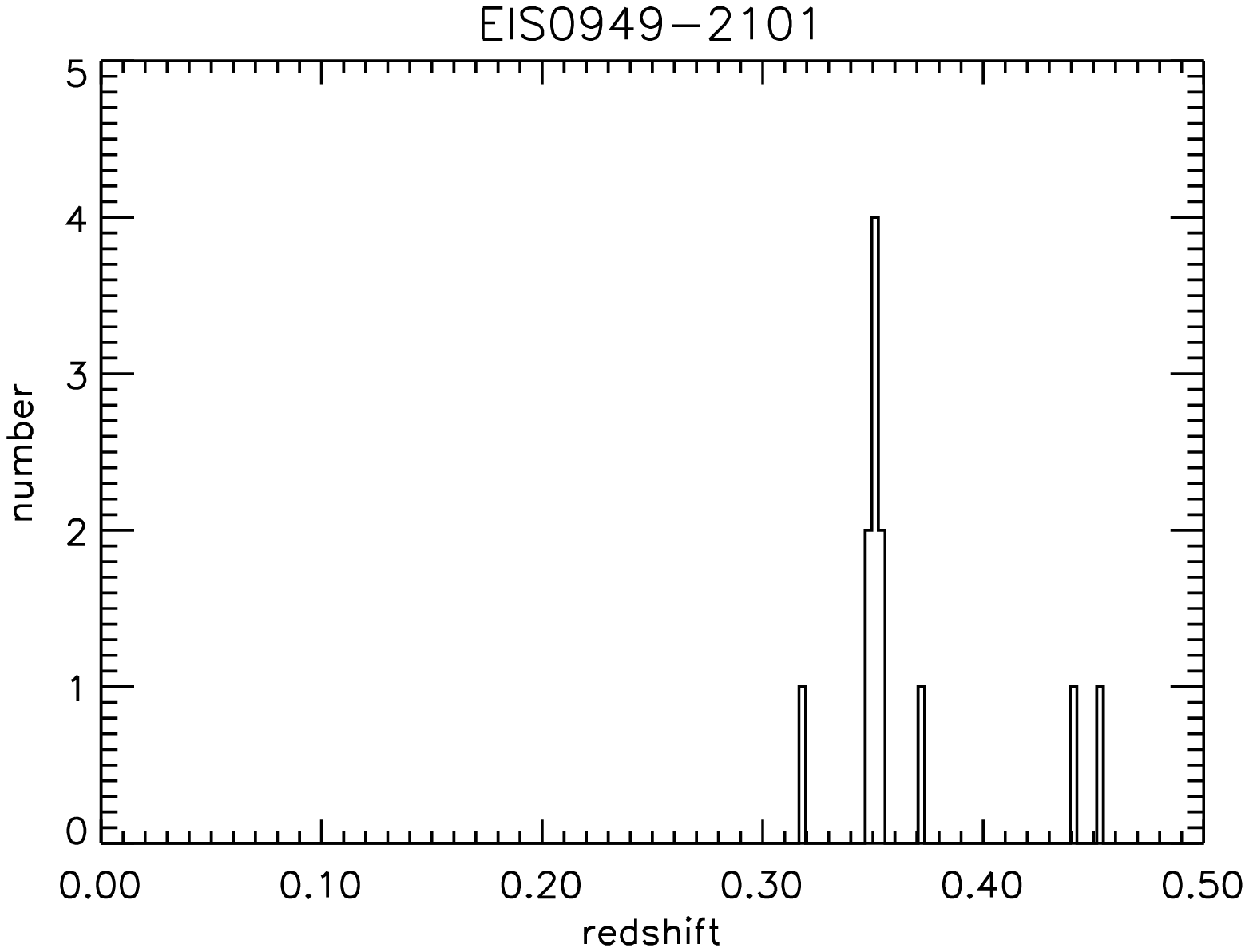}}
  \resizebox{0.5\columnwidth}{!}{\includegraphics{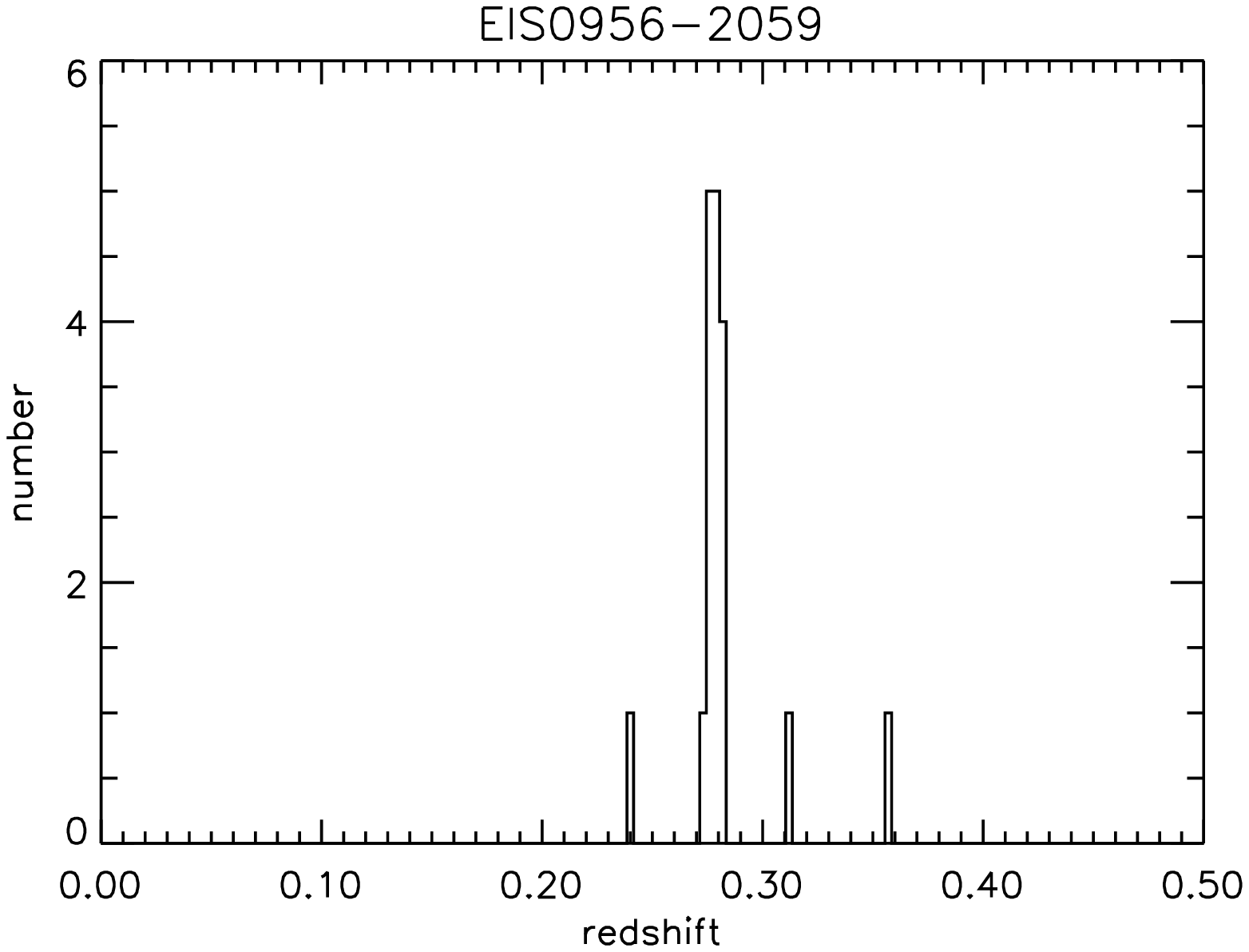}}
  \caption{
     The distribution of redshifts determined in the
     5 cluster fields as indicated in each panel.
        }
  \label{EIS061}
\end{center}
\end{figure*}
%


\section{Data reduction}


The data reduction was performed using the IRAF\footnote{ IRAF is
distributed by the National Optical Astronomy Observatories, which is
operated by AURA Inc. under contract with NSF } package. Bias was
determined from overscan regions and subtracted. After flat fielding
with the dome flats the resulting images were examined for residual
fringing, resulting from shifts of the positions of the slit images
between science exposures and flat fields. These shifts originate in
instrument flexure, that causes the slit images to move slightly with
position on the sky.

The flexure induced image shifts are apparent from exposure to
exposure even near zenith. As mentioned above flat fields had to be
obtained against a screen fixed on the dome, which led to typical
shifts of 2-4 pixels between science exposures and exposures on the
screen. This is unfortunate because the red part of the flat fields 
show high frequency variations caused by fringing in the thinned and 
backside illuminated EEV/MAT chip used. The peak-to-peak distance 
between fringes is only $\approx 15$ pixels. Even small shifts between 
flat field and science exposures will therefore show residual fringing 
in the spectra compromising faint spectra. The present data from the
test observations are, therefore, not of an optimum quality, and all
frames through mask \#14 were abandoned because of a too large shift
of the flat field. The He-Ne exposures on the other hand
were taken before and after the science exposures without changing 
telescope pointing. Relative flexure shifts were therefore small and
could be taken into account. 

It is worth mentioning that since the test observations a flat field
lamp has been installed in the telescope skybaffle besides the He and
Ne lamps, allowing flat fields to be obtained at the same telescope
position as the science frames. This procedure is expected to improve
the quality of the flat fielding for future MOS observations.

Having discarded the frames with strong residual fringing the
remaining frames were stacked using an IRAF script developed for this
purpose.  After applying small shifts to account for the different
positions of the slit images in the science exposures, the script
corrects for cosmic ray events and known defects of the chip.

The individual spectra were extracted and wavelength calibrated using
the standard IRAF tasks. First the image section containing the
two-dimensional spectrum of a slit was copied and wavelength
calibrated by means of the corresponding He-Ne spectrum. Then the
image section was transformed to wavelength scale along the
columns. The background was determined and then subtracted along each
individual line, and a one-dimensional spectrum was finally
extracted. An approximate flux calibration was performed based on
observations of a few spectro\-photometric standards observed through
a $10\,\arcsec$ slit.

Possible residuals from strong sky emission lines were removed by
simple interpolation across the lines. A correction for the
atmospheric B-band absorption around 6870\,\AA~ was performed by
multiplying with a band correction function. This function was
determined from spectro\-photometric standards observed with a $2
\arcsec$ slit. The otherwise smooth spectrum was interpolated across
the band, and the result divided with the original spectrum.

The redshift determination was performed by Fourier cross-correlation
of the spectra with template spectra. The templates were obtained from
Kinney et al. (\cite{Kinney}) and are UV-optical spectra covering from
1200 to 10000\,{\AA}. We used template spectra for elliptical, bulge,
S0, Sa, Sb, and Sc. The Sc template has very strong emission lines and
was suitable for cross correlation with emission line objects.  The
cross-correlation was performed using the IRAF task $fxcor$, and
results are presented in Tables~\ref{EIS0946-2029} through
\ref{EIS0956-2059}.

The templates were always redshifted close to the redshift under
consideration.  Whenever a peak in the correlation function was
accepted as real or possibly real, the observed spectrum was inspected
and compared to the expected positions of the most prominent spectral
features.  We demanded that some features like the H and K lines, the
4000\,{\AA} break, or emission lines should be identified before a
determination was accepted as certain.  This precaution was due to the
possibility that residual fringing could introduce spurious correlation
peaks. If no convincing features were
found, but the correlation peak appeared real, we mark the $z$-value
with a colon in Tables~\ref{EIS0946-2029}-\ref{EIS0956-2059}. If emission
lines are identified the value is marked with an ``e''. In a
number of cases the spectrum suffered from residual fringing or was
too weak to yield a redshift determination.

The accuracy of the measured redshifts are influenced by the limited
resolution, signal-to-noise, fringing, possible systematic errors in
the wavelength transformation, and the consistency with the
template. These error sources vary from spectrum to spectrum. A hint
of the accuracy is given by the typical FWHM $~\Delta z \approx 0.004$
for the Gauss fit to the correlation peak. The position of the peak is
likely to be found with an accuracy $\approx 10\%$ of the FWHM giving an 
expected standard deviation of $\sigma_{z} \approx 0.0004$. 
Another way to estimate the error is by comparing the redshift
estimates obtained for the same galaxy using different template
spectra. From applying all the templates to the same galaxy spectrum
we obtain a standard deviation $\sigma_{z} = 0.0006$ of the derived
redshifts. This is consistent with the error estimated from the width
of the correlation peak. Particularly, since some systematic
differences between the redshifts derived from the different templates
was observed.  E.g. the bulge spectrum has a tendency to give slightly
higher values than the others. All together we estimate an error for
the mean value of $\sigma_{z} \approx 0.0005$.

Only one object has been observed through different masks, namely
EIS0948-2044 \#30 for which spectra are available from masks \#15 and
\#16. The two results deviate by $\sigma_{z} = 0.0006$ consistent with
the above estimate.

%
%
%
%
%
%
%

   \begin{table}
      \caption[]{The total integration time (seconds), number of slits, number 
      of successful redshift determinations, and remarks for the 8 MOS masks }
         \label{plates}
      \center
      \begin{tabular}{l l l r l }
            \hline
 Mask &  exp. & ${\rm N_{slits}}$ & ${\rm N}_{z}$ & remarks \\
            \hline
   11        &      3900   &     19        &    21   &  3 in one slit  \\
   12        &      3600   &     18        &    17   &    \\
   13        &      6000   &     24        &    18   &    \\
   14        &      3600   &     23        &     0   &  no flat field  \\
   15        &      7200   &     24        &    21   &     \\
   16        &      7200   &     22        &    17   &     \\
   17        &      4800   &     20        &    19   &     \\
   18        &      6000   &     25        &    12   &     \\
            \hline
         \end{tabular}
         \end{table}


\section{Results}

Tables~\ref{EIS0946-2029} through \ref{EIS0956-2059} give the measured
redshifts for all the galaxies. The positions and photometry are from 
Benoist et al. (\cite{Benoist99}) and da~Costa et al. (\cite{daCosta}).
Magnitudes are total magnitudes determined by SExtractor (MAG\_AUTO,
Bertin and Arnouts \cite{BertArn}), and additionally corrected for 
interstellar extinction as described by Olsen~(\cite{Olsen_thesis}).

We have secured between 12 and 38   
redshifts per cluster field. The redshift histograms for each field
are shown in Fig.~\ref{EIS061}. In all fields an
obvious peak in the distribution is found indicating the presence of a
cluster.

The cluster redshifts are determined as the mean value for all objects
in the peak.  Firstly, we identify cluster members by accepting all galaxies
within $\Delta z \le 0.01$ from the mean redshift.  Then
we apply a sigma-clipping at 3.5\,$\sigma$ to discard outliers, and
the redshift of the cluster is determined as the mean of the redshifts
of the remaining member galaxies.  The cluster members are those
galaxies with their redshift printed in boldface in
Tables~\ref{EIS0946-2029}-\ref{EIS0956-2059}. We note the high success
rate in picking out cluster members which is due to our selection of
objects falling on the red sequence.

Table~\ref{Zval} gives the cluster redshifts, $z_{\rm cl}$, the observed 
standard 
deviation in redshift among the cluster members, $\sigma_{\rm cl}$, the 
line-of-sight velocity dispersion, ${\rm V_{r}}$, and the number of member 
galaxies
found in our samples.  The measured cluster redshifts are in good agreement
with the matched filter estimated values ($ z_{\rm mf}$ in 
Table~\ref{ObsClusters}).  ${\rm V_{r}}$ is derived from $\sigma_{\rm cl}$ 
after 
correction for an assumed measuring uncertainty $\sigma_z=0.0005$ and has been 
transferred to the cluster rest frame by division with $1+z$. We find
line-of-sight velocity dispersions in the range 294--621\,km/s, which is
comparable to that of relatively poor clusters of galaxies in the Abell
catalog and in agreement with the estimated
richness classes of the EIS cluster candidates of R=0--1 (Olsen
\cite{Olsen_thesis}). 

In three fields secondary peaks due to possible background
concentrations are identified. These are in the fields of EIS0946-2029
(4 objects, $z$=0.1361$\pm$0.0002), EIS0947-2120 (5 objects,
$z$=0.2215$\pm$0.0003), and EIS0948-2044 (3 objects,
$z$=0.2671$\pm$0.0004; 2 objects, $z$=0.3510$\pm$0.0002). Assuming
that these concentrations are bound systems the quoted errors can be used to
derive an upper limit on the accuracy of the individual measurements,
by adopting a negligible velocity dispersion of the systems. From these
four systems we estimate the error of an individual redshift
measurement to be in the range $\sigma_z=0.0003-0.0008$, which
supports the previous estimated accuracy of $\sigma_{z} \approx
0.0005$.

   \begin{table}
      \caption[]{ Observed cluster redshift, observed dispersion in redshift,
         corrected line-of-sight velocity dispersion (km/sec),
         and the number of observed members }
         \label{Zval}
      \begin{tabular}{l c c c c }
            \hline
 Cluster & $z_{\rm cl}$  & $\sigma_{\rm cl}$ & ${\rm V_{r}}$ & objects  \\
            \hline
 EIS0946-2029 & 0.1107 & 0.00172 & 445 & 28  \\
 EIS0947-2120 & 0.1910 & 0.00127 & 294 & 12  \\
 EIS0948-2044 & 0.1818 & 0.00168 & 407 & 27  \\
 EIS0949-2101 & 0.3528 & 0.00247 & 537 & 08   \\
 EIS0956-2059 & 0.2797 & 0.00270 & 621 & 15   \\
            \hline
         \end{tabular}
         \end{table}


\section{Conclusions}

In this work we demonstrate that MOS is possible at the Danish 1.54m
telescope  notwithstanding the high mechanical requirements due to the 
small image scale.  Flat fielding of MOS observations was not optimum during
the test observations, but subsequent installation of a flat field
lamp in the skybaffle allows future flat fields to be obtained at the same
telescope position as the science exposure. The new procedure is
expected to yield flat
fielding of good quality.  In spite of the flat field problems we were
able to determine redshifts for 124 galaxies from the test data. We
estimate the accuracy of the measured redshifts to be $\sigma_{z}
\approx 0.0005$.

The test observations were targeting the fields of 5 EIS cluster
candidates selected for their prominent red sequences.  Multi-slit
masks were constructed to cover members of the red sequences, which
leads to a high success rate in picking out cluster members, yielding
well-defined cluster redshifts as well as velocity dispersions. 

In addition to  determining cluster redshifts the data can 
be used to investigate the quality of the cluster characteristics  
listed in the EIS Cluster Candidate Catalog (Olsen et al., \cite{Olsen99a}; 
\cite{Olsen99b}; Scodeggio et al., \cite{Scodeggio99}). Firstly, we find 
that the measured redshifts are in good agreement with those
originally estimated from the matched filter procedure. Secondly, we note 
that the derived
velocity dispersions are consistent with the clusters being of Abell
richness class R=0--1 as estimated for the EIS cluster candidates.
Both of these results show that the EIS clusters are well-described by their
characteristics in the EIS Cluster Candidate Catalog.

The results obtained in this work indicate that MOS on clusters with
redshifts up to $z \approx 0.4$ is within reach with the Danish
1.54\,m telescope at La~Silla. Furthermore, the confirmation of 5 EIS
clusters strongly supports the reality of the EIS cluster
candidates. The observations reported here are the first in a
systematic program of spectroscopic investigations of the low-redshift
sample of the EIS cluster candidates.


\begin{acknowledgements}
We thank John Pritchard for making the pre-imaging observations of the
fields. We would also like to thank the 2p2 team, La Silla, for their
support, particularly Emanuela Pompei for her important help in the
development of the alignment software.  We are in debt to Michael
I. Andersen, Per Kj{\ae}rgaard, Morten Liborius Jensen, and
Anton Norup S{\o}rensen for their great work in developing the MOS
mode for DFOSC.  We thank the referee Dr. P. Katgert for useful
comments. This work has been supported by The Danish Board for
Astronomical Research.  LFO thanks the Carlsberg Foundation for
financial support.
\end{acknowledgements}

\begin{appendix}

\section{Measured redshifts}

%
   \begin{table}
      \caption[]{Redshifts obtained in the EIS0946-2029 field.
        A cluster redshift $z_{\rm cl}= 0.1107$ is determined. Redshifts 
        for accepted members, i.e.
        deviating less than $3.5 \times \sigma_{\rm cl}$  (Table~~\ref{Zval}), 
        are printed in boldface. Whenever no convincing spectral features
        could be identified despite of a clear correlation peak the value
        is followed by a colon. An ``e'' marks objects with identified
        emission lines }
        \label{EIS0946-2029}
      \begin{tabular}{r c c c c c }
            \hline
   &${\rm \alpha_{J2000}}$ & ${\rm \delta_{J2000}}$ & V & V-I & $z$  \\ 
            \hline
 1 & 09:45:43.9 & -20:31:51 &  19.73 &  1.12& {\bf 0.1105:} \\
 2 & 09:45:46.2 & -20:31:43 &  18.74 &  1.08& {\bf 0.1106~} \\
 3 & 09:45:47.4 & -20:33:17 &  20.96 &  1.35& {    0.1975~} \\
 4 & 09:45:49.4 & -20:30:27 &  19.42 &  1.14& {    0.2180~} \\
 5 & 09:45:51.1 & -20:28:03 &  18.82 &  1.14& {\bf 0.1105~} \\
 6 & 09:45:51.7 & -20:32:58 &  17.59 &  1.24& {\bf 0.1136e} \\
 7 & 09:45:55.2 & -20:28:21 &  18.19 &  1.20& {    0.0000~} \\
 8 & 09:46:00.4 & -20:33:32 &  19.58 &  1.12& {\bf 0.1114~} \\
 9 & 09:46:00.7 & -20:32:56 &  18.13 &  1.28& {\bf 0.1103~} \\
10 & 09:46:02.1 & -20:32:38 &  19.71 &  1.15& {\bf 0.1123~} \\
11 & 09:46:03.0 & -20:31:12 &  17.09 &  1.26& {\bf 0.1073~} \\
12 & 09:46:04.3 & -20:33:22 &  20.58 &  1.21& {    0.3370:} \\
13 & 09:46:05.3 & -20:29:14 &  18.44 &  1.34& {\bf 0.1114~} \\
14 & 09:46:05.8 & -20:32:35 &  19.48 &  1.16& {\bf 0.1128~} \\
15 & 09:46:07.8 & -20:33:47 &  18.27 &  1.21& {    0.1359e} \\
16 & 09:46:10.5 & -20:29:27 &  18.28 &  1.24& {\bf 0.1120~} \\
17 & 09:46:11.5 & -20:29:26 &  17.14 &  1.25& {\bf 0.1112~} \\
18 & 09:46:11.9 & -20:31:40 &  19.49 &  1.27& {\bf 0.1112~} \\
19 & 09:46:12.4 & -20:29:28 &  17.55 &  1.27& {\bf 0.1119~} \\
20 & 09:46:13.8 & -20:29:53 &  19.23 &  1.32& {    0.3550~} \\
21 & 09:46:14.3 & -20:28:35 &  17.90 &  1.26& {\bf 0.1124~} \\
22 & 09:46:16.2 & -20:33:55 &  18.53 &  1.24& {\bf 0.1082~} \\
23 & 09:46:18.0 & -20:26:59 &  19.57 &  1.34& {\bf 0.1085~} \\
24 & 09:46:18.8 & -20:30:35 &  17.99 &  1.12& {\bf 0.1084~} \\
25 & 09:46:21.5 & -20:33:51 &  18.57 &  1.19& {\bf 0.1109~} \\
26 & 09:46:22.1 & -20:27:45 &  19.56 &  1.17& {\bf 0.1100~} \\
27 & 09:46:22.5 & -20:28:11 &  17.88 &  1.21& {\bf 0.1098~} \\
28 & 09:46:24.3 & -20:31:06 &  17.47 &  1.23& {\bf 0.1081e} \\
29 & 09:46:24.5 & -20:31:19 &  19.08 &  1.17& {\bf 0.1112~} \\
30 & 09:46:25.4 & -20:28:40 &  19.64 &  1.14& {\bf 0.1139~} \\
31 & 09:46:26.2 & -20:26:43 &  18.47 &  1.24& {\bf 0.1078~} \\
32 & 09:46:28.4 & -20:33:56 &  20.27 &  1.18& {    0.1357:} \\
33 & 09:46:29.3 & -20:33:32 &  18.68 &  1.12& {\bf 0.1111~} \\
34 & 09:46:29.9 & -20:33:52 &  18.60 &  1.13& {    0.1362e} \\
35 & 09:46:30.7 & -20:26:17 &  18.88 &  1.33& {    0.0980e} \\
36 & 09:46:32.6 & -20:33:29 &  17.40 &  1.07& {    0.1365~} \\
37 & 09:46:33.1 & -20:29:24 &  19.36 &  1.09& {\bf 0.1105~} \\
38 & 09:46:36.7 & -20:28:18 &  16.55 &  1.10& {\bf 0.1121~} \\
            \hline
         \end{tabular}
         \end{table}
   \begin{table}
      \caption[]{Redshifts obtained in the EIS0947-2120 field.  \\
          A cluster redshift of $z_{\rm cl}= 0.1910$ is determined }
         \label{EIS0947-2120}
      \begin{tabular}{r c c c c c }
            \hline
   &${\rm \alpha_{J2000}}$ & ${\rm \delta_{J2000}}$ & V & V-I & $z$  \\
            \hline
 1 & 09:46:43.2 & -21:20:34 &  19.61 &  1.33& {\bf 0.1919:} \\
 2 & 09:46:46.6 & -21:21:26 &  19.25 &  1.39& {\bf 0.1912~} \\
 3 & 09:46:51.8 & -21:20:18 &  19.00 &  1.47& {\bf 0.1911~} \\
 4 & 09:46:53.0 & -21:24:54 &  18.33 &  1.44& {    0.1402~} \\
 5 & 09:46:53.4 & -21:18:27 &  20.31 &  1.46& {\bf 0.1895:} \\
 6 & 09:46:57.1 & -21:21:04 &  19.01 &  1.46& {\bf 0.1925~} \\
 7 & 09:46:58.1 & -21:21:39 &  19.40 &  1.43& {\bf 0.1912~} \\
 8 & 09:47:00.3 & -21:23:06 &  19.04 &  1.52& {\bf 0.1889~} \\
 9 & 09:47:01.7 & -21:22:26 &  19.45 &  1.60& {    0.2205~} \\
10 & 09:47:03.2 & -21:20:54 &  18.65 &  1.43& {\bf 0.1906~} \\
11 & 09:47:04.6 & -21:19:20 &  19.41 &  1.43& {\bf 0.1896~} \\
12 & 09:47:05.9 & -21:20:08 &  17.65 &  1.42& {\bf 0.1911~} \\
13 & 09:47:07.2 & -21:21:39 &  18.90 &  1.56& {\bf 0.1934~} \\
14 & 09:47:10.6 & -21:23:53 &  19.56 &  1.51& {\bf 0.1907~} \\
15 & 09:47:25.8 & -21:20:11 &  19.92 &  1.54& {    0.2214:} \\
16 & 09:47:28.4 & -21:22:54 &  18.98 &  1.52& {    0.2226~} \\
17 & 09:47:29.9 & -21:22:29 &  19.94 &  1.59& {    0.2215~} \\
18 & 09:47:35.4 & -21:23:38 &  18.67 &  1.47& {    0.2215~} \\
            \hline
         \end{tabular}
         \end{table}
   \begin{table}
      \caption[]{Redshifts obtained in the EIS0948-2044 field. \\ 
          A cluster redshift of $z_{\rm cl}= 0.1818$ is determined}
         \label{EIS0948-2044}
      \begin{tabular}{r c c c c c }
            \hline
   &${\rm \alpha_{J2000}}$ & ${\rm \delta_{J2000}}$ & V & V-I & $z$  \\ 
          \hline
 1 & 09:47:45.3 & -20:41:41 &  18.31 &  1.50& {\bf 0.1826~} \\
 2 & 09:47:45.3 & -20:42:24 &  20.20 &  1.33& {    0.2497:} \\
 3 & 09:47:52.8 & -20:46:36 &  20.37 &  1.38& {\bf 0.1816:} \\
 4 & 09:47:54.2 & -20:41:23 &  18.59 &  1.43& {\bf 0.1805~} \\
 5 & 09:47:55.6 & -20:43:49 &  19.48 &  1.44& {    0.2225~} \\
 6 & 09:47:56.4 & -20:44:21 &  19.21 &  1.40& {\bf 0.1838~} \\
 7 & 09:47:56.5 & -20:40:47 &  18.09 &  1.49& {\bf 0.1818~} \\
 8 & 09:47:56.8 & -20:41:53 &  20.84 &  1.42& {\bf 0.1800~} \\
 9 & 09:47:57.8 & -20:42:39 &  19.34 &  1.47& {\bf 0.1842~} \\
10 & 09:47:58.9 & -20:44:18 &  21.20 &  1.27& {\bf 0.1792~} \\
11 & 09:47:59.6 & -20:42:13 &  19.89 &  1.39& {\bf 0.1813~} \\
12 & 09:48:00.5 & -20:43:14 &  19.82 &  1.36& {\bf 0.1790~} \\
13 & 09:48:03.6 & -20:45:51 &  19.11 &  1.35& {\bf 0.1847~} \\
14 & 09:48:05.2 & -20:43:46 &  18.04 &  1.50& {\bf 0.1830~} \\
15 & 09:48:05.2 & -20:44:32 &  19.44 &  1.48& {\bf 0.1824~} \\
16 & 09:48:05.8 & -20:43:57 &  19.69 &  1.39& {\bf 0.1812~} \\
17 & 09:48:06.5 & -20:42:53 &  19.63 &  1.39& {\bf 0.1871~} \\
18 & 09:48:08.6 & -20:44:03 &  19.14 &  1.39& {\bf 0.1833~} \\
19 & 09:48:08.7 & -20:45:00 &  19.68 &  1.40& {\bf 0.1817~} \\
20 & 09:48:09.9 & -20:43:15 &  20.63 &  1.49& {    0.2158~} \\
21 & 09:48:10.4 & -20:45:22 &  19.04 &  1.51& {\bf 0.1821~} \\
22 & 09:48:11.1 & -20:45:30 &  19.49 &  1.54& {    0.3512e} \\
23 & 09:48:11.4 & -20:45:33 &  20.29 &  1.41& {    0.3508e} \\
24 & 09:48:12.4 & -20:48:49 &  19.31 &  1.33& {\bf 0.1830:} \\
25 & 09:48:12.4 & -20:43:14 &  21.02 &  1.33& {    0.1372:} \\
26 & 09:48:14.4 & -20:46:48 &  19.60 &  1.34& {\bf 0.1843~} \\
27 & 09:48:14.8 & -20:45:23 &  18.90 &  1.44& {\bf 0.1835~} \\
28 & 09:48:16.5 & -20:44:32 &  19.73 &  1.42& {\bf 0.1814~} \\
29 & 09:48:17.1 & -20:42:28 &  18.51 &  1.26& {    0.1107~} \\
30 & 09:48:20.7 & -20:45:16 &  20.29 &  1.53& {    0.2662~} \\
31 & 09:48:22.1 & -20:48:13 &  18.29 &  1.26& {\bf 0.1798~} \\
32 & 09:48:23.7 & -20:44:32 &  20.87 &  1.39& {\bf 0.1816~} \\
33 & 09:48:24.5 & -20:44:38 &  20.02 &  1.55& {    0.2676~} \\
34 & 09:48:24.8 & -20:45:24 &  20.63 &  1.26& {\bf 0.1798~} \\
35 & 09:48:26.2 & -20:48:03 &  19.88 &  1.41& {\bf 0.1791e} \\
36 & 09:48:30.9 & -20:43:17 &  20.42 &  1.28& {\bf 0.1830:} \\
37 & 09:48:35.6 & -20:45:39 &  20.34 &  1.27& {    0.2674:} \\
            \hline
         \end{tabular}
         \end{table}
   \begin{table}
      \caption[]{Redshifts obtained in the EIS0949-2101 field. \\
          A cluster redshift of $z_{\rm cl}= 0.3528$ is determined}
         \label{EIS0949-2101}
      \begin{tabular}{r c c c c c }
            \hline
   &${\rm \alpha_{J2000}}$ & ${\rm \delta_{J2000}}$ & V & V-I & $z$  \\ 
            \hline
 1 & 09:49:02.2 & -21:00:33 &  21.36 &  2.14& {    0.4418:} \\
 2 & 09:49:05.0 & -21:02:59 &  20.38 &  2.17& {    0.4536e} \\
 3 & 09:49:12.2 & -20:59:05 &  21.84 &  2.05& {    0.3730:} \\
 4 & 09:49:13.4 & -20:59:25 &  21.69 &  2.13& {\bf 0.3500~} \\
 5 & 09:49:15.8 & -20:59:52 &  20.92 &  2.03& {\bf 0.3499~} \\
 6 & 09:49:19.0 & -21:02:08 &  21.30 &  2.05& {\bf 0.3539~} \\
 7 & 09:49:20.0 & -21:01:07 &  20.66 &  2.02& {\bf 0.3510~} \\
 8 & 09:49:20.6 & -21:01:16 &  19.78 &  1.98& {\bf 0.3535~} \\
 9 & 09:49:23.1 & -21:02:05 &  20.32 &  2.10& {\bf 0.3520~} \\
10 & 09:49:25.1 & -21:02:19 &  20.69 &  2.01& {\bf 0.3555~} \\
11 & 09:49:32.0 & -21:01:18 &  21.06 &  2.08& {\bf 0.3565~} \\
12 & 09:49:42.0 & -21:04:13 &  21.52 &  1.92& {    0.3181~} \\
            \hline
         \end{tabular}
         \end{table}
   \begin{table}
      \caption[]{Redshifts obtained in the EIS0956-2059 field. \\ 
          A cluster redshift of $z_{\rm cl}=0.2797$ is determined}
         \label{EIS0956-2059}
      \begin{tabular}{r c c c c c }
            \hline
   &${\rm \alpha_{J2000}}$ & ${\rm \delta_{J2000}}$ & V & V-I & $z$  \\ 
          \hline
 1 & 09:56:12.6 & -20:59:21 &  17.23 &  1.62& {    0.0000~} \\
 2 & 09:56:12.8 & -20:57:10 &  20.54 &  1.77& {\bf 0.2831~} \\
 3 & 09:56:15.8 & -21:02:14 &  20.53 &  1.68& {\bf 0.2804~} \\
 4 & 09:56:17.8 & -20:56:44 &  20.07 &  1.84& {    0.3589~} \\
 5 & 09:56:18.9 & -20:58:55 &  19.93 &  1.62& {    0.3136:} \\
 6 & 09:56:20.6 & -20:57:36 &  20.37 &  1.76& {\bf 0.2787~} \\
 7 & 09:56:24.7 & -20:59:22 &  19.25 &  1.65& {\bf 0.2806~} \\
 8 & 09:56:25.8 & -21:00:40 &  20.08 &  1.79& {\bf 0.2804~} \\
 9 & 09:56:26.6 & -21:00:16 &  19.90 &  1.75& {\bf 0.2784~} \\
10 & 09:56:27.5 & -21:00:36 &  20.58 &  1.70& {\bf 0.2771~} \\
11 & 09:56:28.2 & -20:58:58 &  20.55 &  1.71& {\bf 0.2731~} \\
12 & 09:56:30.1 & -21:01:39 &  20.42 &  1.65& {\bf 0.2837~} \\
13 & 09:56:30.9 & -20:57:48 &  19.58 &  1.65& {    0.2423~} \\
14 & 09:56:32.5 & -21:03:16 &  19.90 &  1.75& {\bf 0.2822~} \\
15 & 09:56:34.3 & -20:58:44 &  20.36 &  1.68& {\bf 0.2821~} \\
16 & 09:56:37.0 & -20:57:57 &  19.89 &  1.56& {\bf 0.2775~} \\
17 & 09:56:37.9 & -21:02:13 &  19.19 &  1.71& {\bf 0.2805~} \\
18 & 09:56:41.8 & -21:03:18 &  19.59 &  1.69& {\bf 0.2783~} \\
19 & 09:56:47.3 & -20:57:38 &  20.48 &  1.81& {\bf 0.2796~} \\
            \hline
         \end{tabular}
         \end{table}

%

\end{appendix}


\begin{thebibliography}{}

  \bibitem[1999]{Benoist99} Benoist C., da~Costa L., Olsen L.F., et al.
      1999, A\&A, 346, 58

  \bibitem[2001]{Benoist} Benoist C., da Costa L., J{\o}rgensen H.E.,
      et al.  2001, preprint to be published in A\&A

  \bibitem[1996]{BertArn} Bertin E., \& Arnouts S. 1996, A\&AS, 117, 393

  \bibitem[2000]{daCosta} da~Costa L., Arnouts S., Benoist C., et al.
      2000, {\it The Messenger}, 98, 36

  \bibitem[1999]{Holden} Holden B.P., Nichol R.C., Romer A.K., et al.
      1999, AJ 118, 2002

  \bibitem[2001]{Jones} Jones H., Pompei E., et al. 2001, {\it The Messenger} 
      104, 7

  \bibitem[1996]{Kinney} Kinney A.L., Calzetti D., Bohlin R.C., et al.
      1996, ApJ 467, 38

  \bibitem[2002]{Neumann} Neumann D.M., Benoist C., Arnaud M., et al. 
      2002, to appear in Proc. Symposium ``New Visions of the X-ray 
      Universe in the XMM-Newton and Chandra Era''

  \bibitem[1999a]{Olsen99a} Olsen L.F., Scodeggio M., da Costa L., et al.
      1999, A\&A 345, 363 

  \bibitem[1999b]{Olsen99b} Olsen L.F., Scodeggio M., da Costa L., et al.
      1999, A\&A 345, 681 

  \bibitem[2000]{Olsen_thesis} Olsen L.F. 2000, Ph.D.-dissertation

  \bibitem[1996]{Postman96} Postman M, Lubin L.M., Gunn J.E., et al.
      1996, AJ 111, 615

  \bibitem[2000]{Ramella} Ramella M., Biviano A., Boschin W., et al.
      2000, A\&A, 360, 861

  \bibitem[1999]{Scodeggio99} Scodeggio M., Olsen L.F., da Costa L., et al.
      1999, A\&AS 137, 83

\end{thebibliography}
\end{document}